\documentclass[aps,prl,reprint,superscriptaddress]{revtex4-2}
\usepackage{graphicx}

\usepackage{color,amsmath}
\usepackage[normalem]{ulem}
\usepackage{hyperref}
\usepackage{lipsum}
\usepackage{xcolor}
\usepackage{float}
\usepackage{nicefrac}

\makeatletter
\def\@fnsymbol#1{\ensuremath{\ifcase#1\or \ddagger\or *\or \dagger\or	\mathsection\or \mathparagraph\or **\or \dagger\dagger\or \ddagger\ddagger \else\@ctrerr\fi}}
\makeatother

\begin{document}
\title{Large even-odd spacing and $g$-factor anisotropy in PbTe quantum dots}

\author{S.~C.~ten~Kate}
\thanks{S.~C.~ten~Kate and M.~F.~Ritter contributed equally to this paper.}
\affiliation{IBM Research Europe, S\"aumerstrasse 4, 8803 R\"uschlikon, Switzerland}
\affiliation{University of Twente, Drienerlolaan 5, 7522 NB Enschede, Netherlands}

\author{M.~F.~Ritter}
\thanks{S.~C.~ten~Kate and M.~F.~Ritter contributed equally to this paper.}
\affiliation{IBM Research Europe, S\"aumerstrasse 4, 8803 R\"uschlikon, Switzerland}

\author{A.~Fuhrer}
\affiliation{IBM Research Europe, S\"aumerstrasse 4, 8803 R\"uschlikon, Switzerland}

\author{J.~Jung}
\affiliation{Eindhoven University of Technology, 5600 MB Eindhoven, The Netherlands}

\author{S.~G.~Schellingerhout}
\affiliation{Eindhoven University of Technology, 5600 MB Eindhoven, The Netherlands}

\author{E.~P.~A.~M.~Bakkers}
\affiliation{Eindhoven University of Technology, 5600 MB Eindhoven, The Netherlands}

\author{H.~Riel}
\affiliation{IBM Research Europe, S\"aumerstrasse 4, 8803 R\"uschlikon, Switzerland}

\author{F.~Nichele}
\email[email: ]{fni@zurich.ibm.com}
\affiliation{IBM Research Europe, S\"aumerstrasse 4, 8803 R\"uschlikon, Switzerland}

\date{\today}

\begin{abstract}
PbTe is a semiconductor with promising properties for topological quantum computing applications. Here we characterize quantum dots in PbTe nanowires selectively grown on InP. Charge stability diagrams at zero magnetic field reveal large even-odd spacing between Coulomb blockade peaks, charging energies below 140~\textmu eV and Kondo peaks in odd Coulomb diamonds. We attribute the large even-odd spacing to the large dielectric constant and small effective electron mass of PbTe. By studying the Zeeman-induced level and Kondo splitting in finite magnetic fields, we extract the electron $g$-factor as a function of magnetic field direction. We find the $g$-factor tensor to be highly anisotropic, with principal $g$-factors ranging from 0.9 to 22.4, and to depend on the electronic configuration of the devices. These results indicate strong Rashba spin-orbit interaction in our PbTe quantum dots.
\end{abstract}
\maketitle

The quest for realizing topological superconductivity in trivial semiconductors, with accompanying Majorana zero modes, would benefit from materials with strong spin-orbit interaction and large Land\'{e} $g$-factors \cite{Prada2020,Lutchyn2018,Kanne2021,Pendharker2021,OpHetVeld2021}. In this context, PbTe may offer advantages compared to more established platforms such as InSb and InAs. Work on PbTe reported large and anisotropic $g$-factors, with absolute values up to 58 \cite{Patel1969}, strong spin-orbit interaction (SOI) \cite{Peres2011} and small effective masses \cite{Guenther2005}; all advantageous properties for the realization of sizeable topological gaps at moderate magnetic fields \cite{Lutchyn2010,Oreg2010}. PbTe also exhibits a large dielectric constant $\epsilon_{\mathrm{r}}\sim1350$ at low temperatures \cite{Guenther2005} (compared to  $\epsilon_{\mathrm{r}}\sim14$ for InAs and InSb \cite{Rodilla2009}), which is expected to result in efficient screening of impurities and, consequently, high electron mobilities \cite{Jiang2022}. Recent work demonstrated the possibility to grow high-quality PbTe nanowires, either with vapor-liquid-solid epitaxy \cite{Schellingerhout2022} or the selective-area-growth (SAG) technique \cite{Jiang2022}. Electrical characterization also demonstrated ambipolar characteristics, small charging energies and large $g$-factors \cite{Gomanko2021}.

Here, we investigate quantum dots in PbTe nanowires selectively grown on insulating InP substrates. We find that charging energies are typically smaller than single-particle excitation energies, producing a pronounced even-odd spacing between Coulomb blockade peaks that is lifted by applying modest magnetic fields. Such even-odd spacing is consistent with the strong screening expected from the PbTe material. Studying the evolution of spin excited state level splittings and Kondo peaks in a magnetic field, we extract the three-dimensional effective $g$-factor tensor, that is the electronic $g$-factor as a function of magnetic field direction.\\
Our results indicate that the effective $g$-factor tensor is highly anisotropic, moreover it varies from device to device and depends on the gate voltage configuration of each device. In the regimes we investigated, the principal $g$-factors varied from 0.9 to 22.4, with smaller values obtained for magnetic fields parallel to the substrate. No relation between effective $g$-factor tensor and crystal direction was found.

\begin{figure*}
	\includegraphics[width=2\columnwidth]{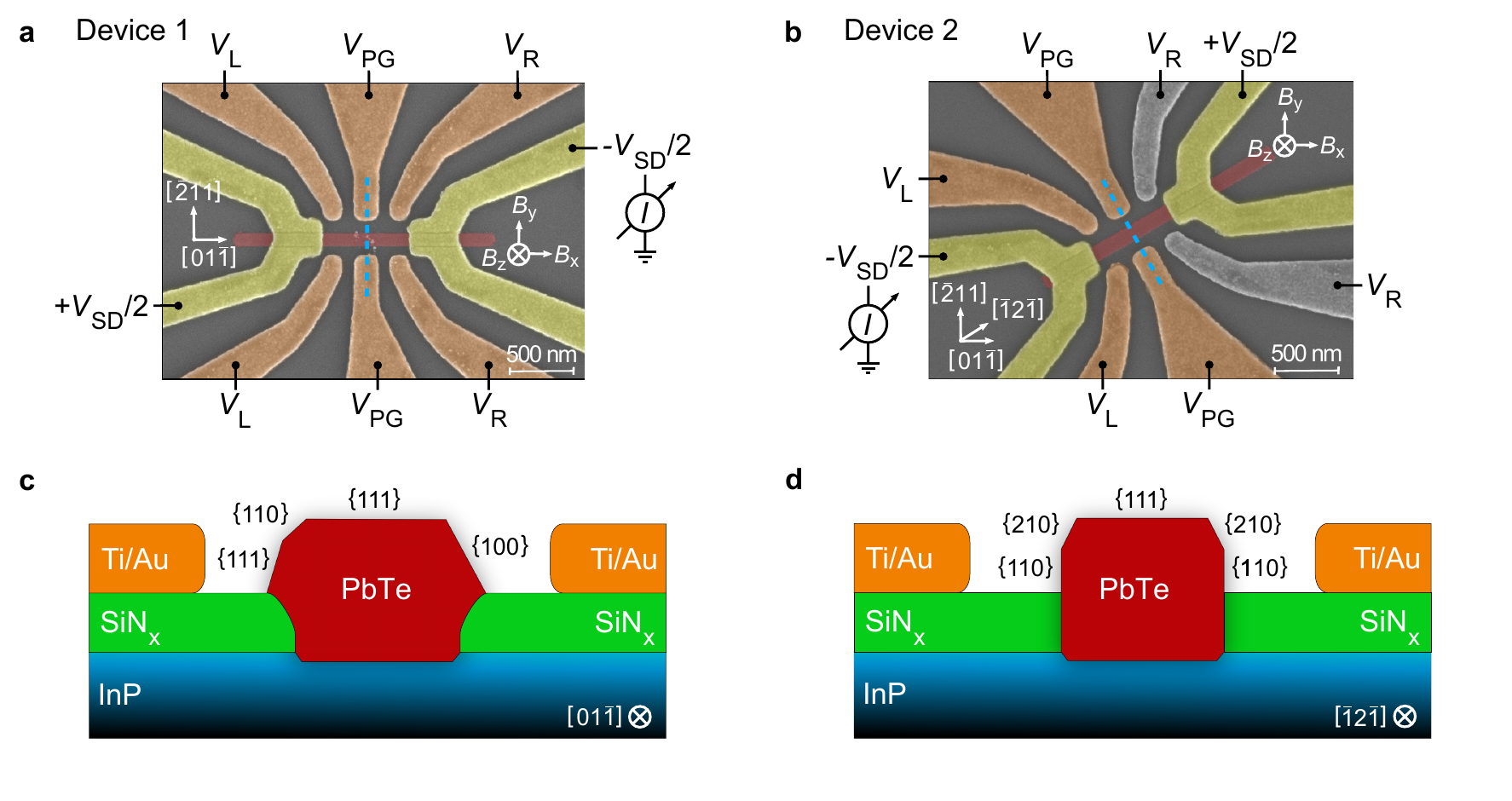}
	\caption{\textbf{Two quantum dot devices in SAG PbTe nanowires on (111)A InP.}
		\textbf{a,b} False-colored SEM micrographs of devices with crystal and magnetic field directions indicated. The nanowires are red, the Ti/Au contacts are yellow, and the Ti/Au gates are orange. For Device~2, $V_\mathrm{R}$ was grounded. \textbf{c,d} Schematic cross sections as indicated in (a) and (b) by blue dashed lines. The terminating facets of the nanowires differ due to their different crystal directions.}
	\label{fig:1}
\end{figure*}

Figure~\ref{fig:1} shows false-colored scanning electron micrographs of the two quantum dot devices used in this study, together with the measurement configurations. The PbTe nanowires are colored red, the Ti/Au contacts yellow and the Ti/Au gates orange. Nanowires were grown in an MBE on a (111)A InP substrate along a $\langle110\rangle$ (Device~1) and a $\langle112\rangle$ (Device~2) crystal direction. The lithographic distance between the source and drain contacts of both quantum dot devices is 720~nm and the width is 80~nm and 100~nm for Device~1 and 2, respectively. Schematics of the device cross-sections in Fig.~\ref{fig:1}(c) and (d) show the InP substrate, SiN${}_\mathrm{x}$ growth mask, PbTe nanowire with terminating facets and Ti/Au side gates. The nanowire cross-section, obtained by TEM imaging of similar nanowires, is a consequence of the crystal direction of the growth mask relative to the substrate and will be discussed in more detail in a separate work.
\\
Measurements were carried out in a dilution refrigerator equipped with a vector magnet at a mixing chamber base temperature below $20~\mathrm{mK}$. A variable DC voltage bias $\pm V_{\mathrm{SD}}/2$ was applied anti-symmetrically to source and drain contacts, superimposed on an AC voltage bias of $3~$\textmu V. The resulting AC current and voltage drop were measured with lock-in amplifiers to determine the differential conductance $G$ of the devices. Both devices were tuned with side gate voltages $V_\mathrm{L}$, $V_{\mathrm{PG}}$, and $V_\mathrm{R}$, applied pairwise to opposite facing gate electrodes. For Device~2, the gray gate pair in Fig.~\ref{fig:1}(b) showed leakage for $V_\mathrm{R} < -600~$mV and was grounded throughout the measurements. In this case, the quantum dot was formed by tuning $V_\mathrm{L}$ and $V_\mathrm{PG}$. Both quantum dots showed gate instabilities over their entire gate voltage space, resulting in frequent charge rearrangements, some of which are visible in the Figures below. The gate voltage regimes characterized in this paper were selected to limit the occurrence of such events.

\begin{figure*}
	\includegraphics[width=2\columnwidth]{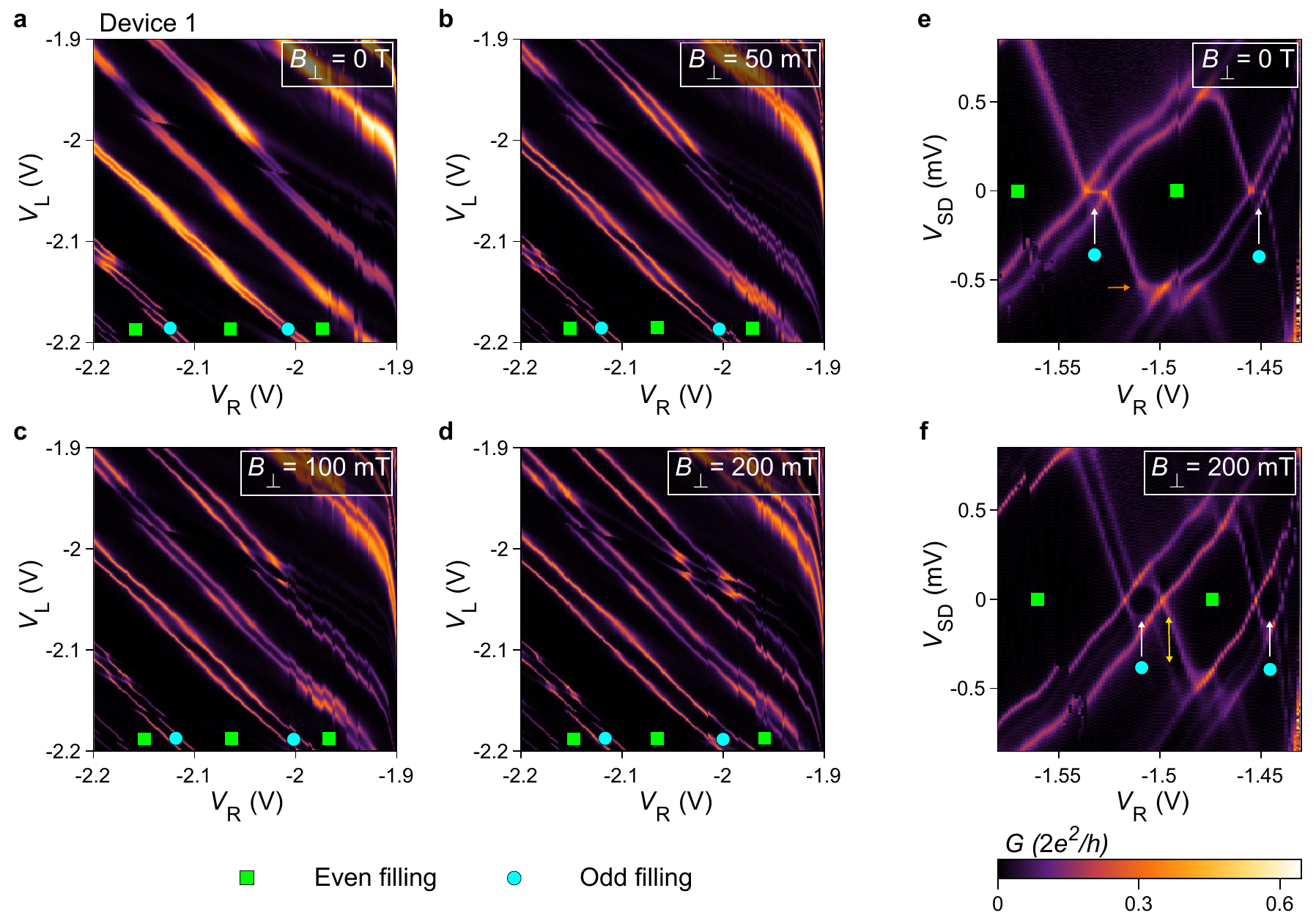}
	\caption{\textbf{Electrical characterization of quantum dot Device~1 at zero and finite magnetic fields.}
		\textbf{a-d} Evolution of even-odd spacing between Coulomb blockade peaks as a function of magnetic field. $V_{\mathrm{PG}}=-1.4~$V is applied to the lower plunger gate in Fig.~\ref{fig:1}(a). \textbf{e-f} Charge stability diagrams showing Kondo peaks, which split in a finite magnetic field. $V_{\mathrm{PG}}=-1.25~$V and $V_{\mathrm{L}}=-2.4~$V. The onset of inelastic cotunneling, which coincides with an excited state, is marked in (e) with an orange arrow and the level splitting is marked in (f) with a yellow double-arrow.}
	\label{fig:2}
\end{figure*}

\begin{figure*}
	\includegraphics[width=2\columnwidth]{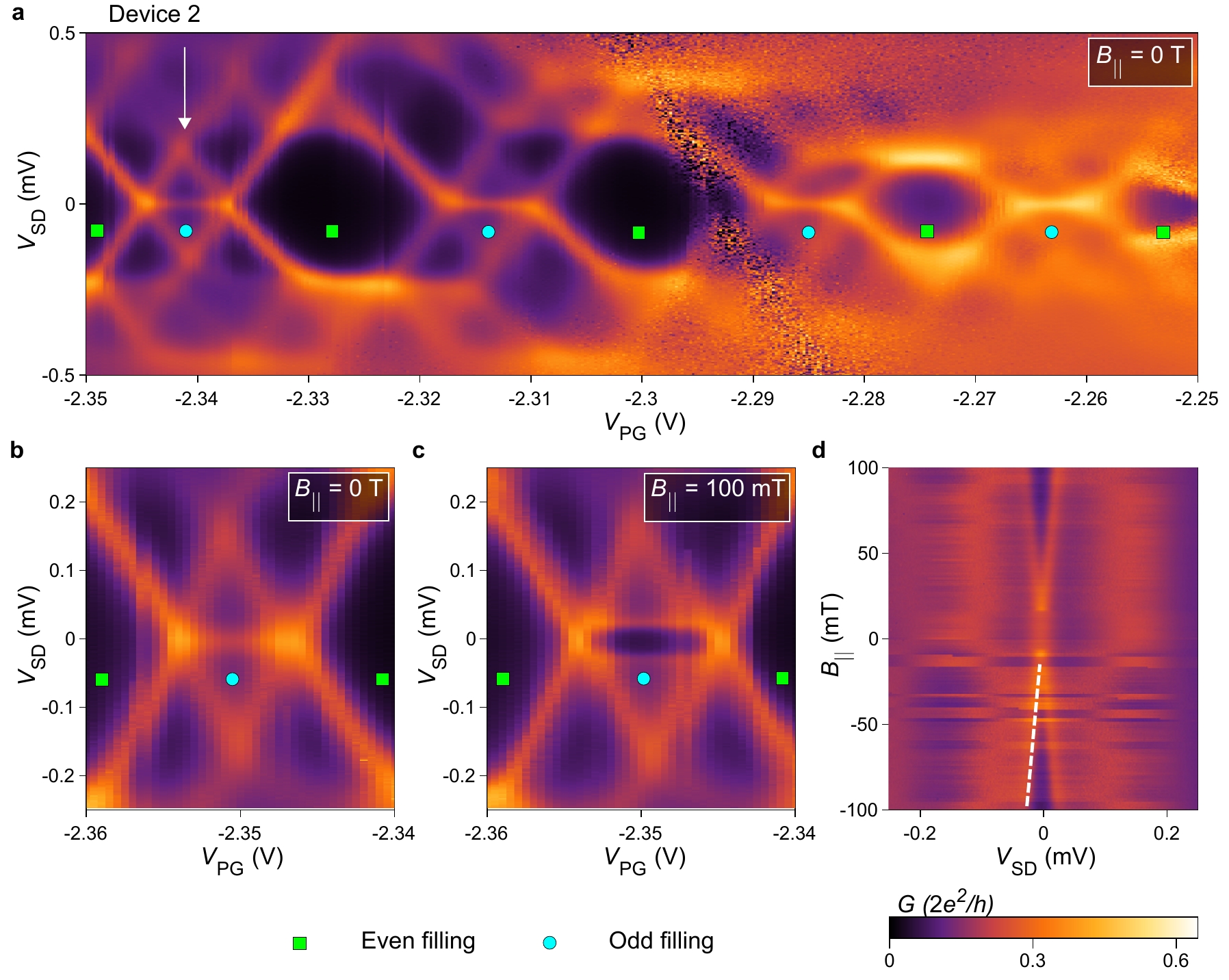}
	\caption{\textbf{Electrical characterization of quantum dot Device~2 at zero and finite magnetic field.}
		\textbf{a} Charge stability diagram at zero magnetic field and $V_\mathrm{L}=-3.825~$V, showing Kondo peaks in odd Coulomb diamonds and inelastic cotunneling in even diamonds. \textbf{b,c} Zoom-ins of the Coulomb diamond indicated with an arrow in (a), depicting the (split) Kondo peak at zero and finite magnetic field. \textbf{d} Evolution of the Kondo peak splitting as a function of magnetic field. The dashed line is a guide for the eye, which shows that the splitting is linear. For a-c: $V_\mathrm{L}=-3.825~$V. For d: $V_\mathrm{L}=-3.825~$V and $V_\mathrm{PG}=-2.344~$V.}
	\label{fig:3}
\end{figure*}

\begin{figure*}
	\includegraphics[width=2\columnwidth]{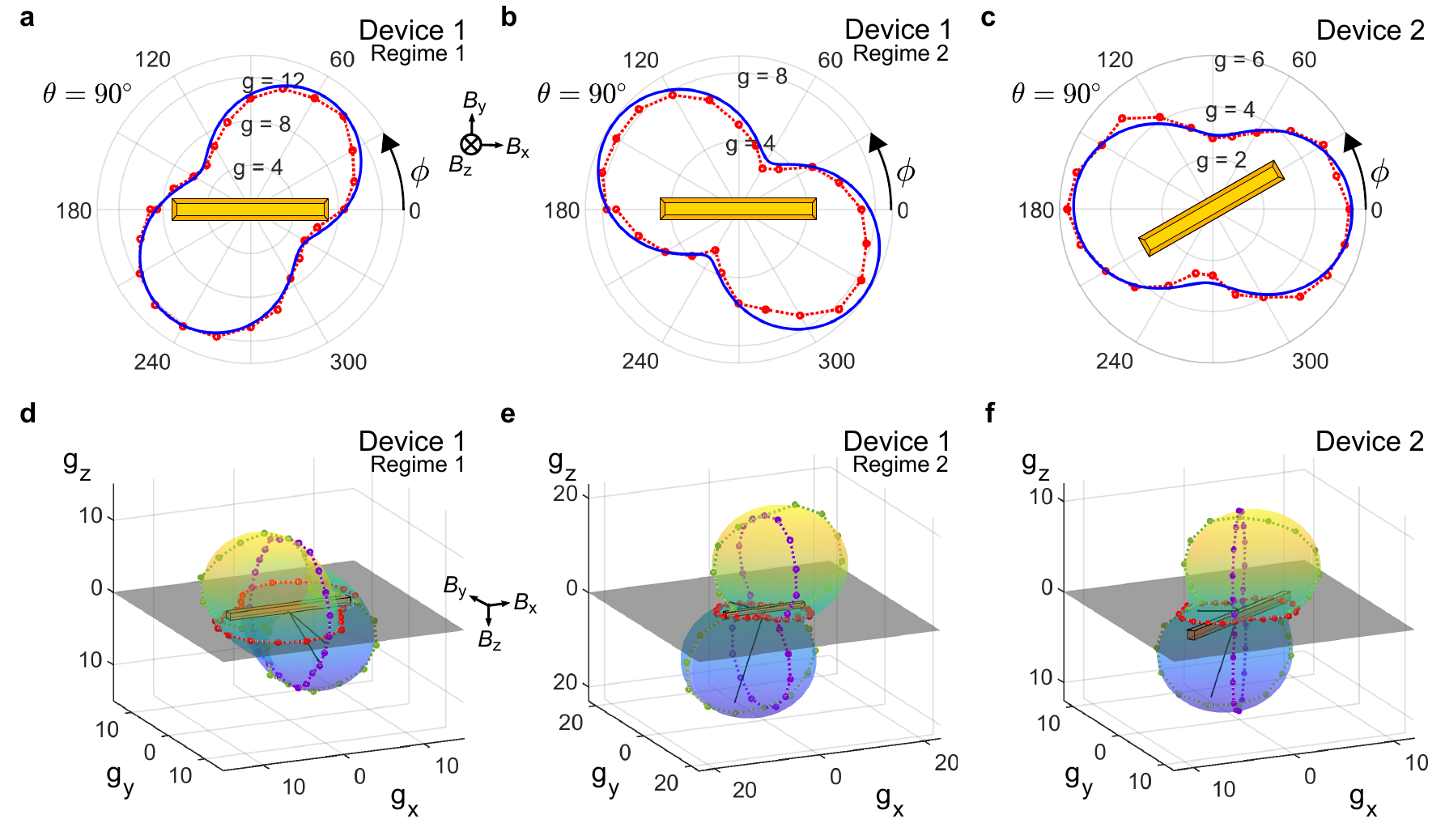}
	\caption{\textbf{$g$-factor anisotropy of all investigated quantum dot regimes.}
		\textbf{a-c} In-plane $g$-factors extracted from energy level and Kondo peak splittings (red) and fits of the effective $g$-factor tensors (blue). The magnetic field was rotated in steps of 15\textdegree. The nanowires are displayed in each polar plot. \textbf{d-f} 3D plots of the $g$-factors extracted from three magnetic field rotations (red, purple, green lines) and the fits of the effective $g$-factor tensors (surface plots and black lines). The nanowire devices (orange), substrate planes (gray) and the magnetic field coordinate system are depicted.}
	\label{fig:4}
\end{figure*}

Our key measurement results are shown in Figs.~\ref{fig:2}, \ref{fig:3}, and \ref{fig:4}, which we will analyze and discuss below. Figure~\ref{fig:2}(a-d) depicts Coulomb blockade measurements as a function of gates $V_\mathrm{L}$ and $V_\mathrm{R}$ for Device~1 in a perpendicular magnetic field. Charge stability diagrams at zero and finite magnetic field are shown in Fig.~\ref{fig:2}(e) and (f), respectively, with $V_{\mathrm{PG}}=-1.25~$V and $V_{\mathrm{L}}=-2.4~$V. The average gate lever arm of gate $V_\mathrm{R}$ was $\alpha_\mathrm{R}=0.0092$. The low value of $\alpha_\mathrm{R}$ is consistent with large source and drain lever arms, accounting for most of the quantum dot capacitance. A discussion of the lever arms can be found in the Supplementary Information.\\ 
The Coulomb peak spacing in Fig.~\ref{fig:2}(a) follows a pronounced even-odd pattern, with the boundaries of even-occupied states (indicated with green squares) much more separated in gate space than those of odd-occupied states (indicated with blue circles). This observation indicates that the charging energy of the quantum dot is much smaller than the orbital energy. At increasing magnetic fields, the closely-spaced Coulomb blockade peaks move further apart, consistent with two electrons of opposite spin filling the same orbital level. From Fig.~\ref{fig:2}(a-d) we verified that the splitting is linear up to 200~mT, at least.\\
The large difference between the charging energy and single-particle excitation energy is evident in Fig.~\ref{fig:2}(e) and (f). The charge stability diagrams of Device~1 show alternating Coulomb diamond sizes, consistent with the large even-odd spacing in Fig.~\ref{fig:2}(a) and (d). From the height of the odd Coulomb diamonds in Fig.~\ref{fig:2}(e), we extracted an average charging energy of $E_\mathrm{C}\approx110~$\textmu eV using $E_{\mathrm{add}}=E_\mathrm{C}$ \cite{Bjork2004}, where $E_{\mathrm{add}}$ is the addition energy. From the height of the central even Coulomb diamond, we extracted a single-particle excitation energy of $\Delta\approx500~$\textmu eV using $E_{\mathrm{add}}=E_\mathrm{C}+\Delta$ \cite{Bjork2004}. Inelastic cotunneling \cite{DeFranceschi2001} is observed near the tips of the even Coulomb diamond and, for the lower tip, the onset of cotunneling coincides with a faint excited state of an odd Coulomb diamond  [see the orange arrow in Fig.~\ref{fig:2}(e)]. In addition, conductance peaks at zero bias voltage are observed in odd Coulomb diamonds, which split in a finite magnetic field perpendicular to the substrate, as seen in Fig.~\ref{fig:2}(f). Therefore, we conclude that the peaks are manifestations of the spin-\nicefrac{1}{2} Kondo effect \cite{GoldhaberGordon1998,Cronenwett1998,Inoshita1998,Jespersen2006}. Charge stability diagrams obtained in a second regime of Device~1 are presented in Supplementary Figure~\ref{fig:S1}, and show similar results.

A charge stability diagram of Device~2 is shown in Fig.~\ref{fig:3}(a). As for Device~1, odd Coulomb diamonds are smaller than even Coulomb diamonds. From the two leftmost odd Coulomb diamonds, we extracted an average charging energy of $E_\mathrm{C}\approx 130~$\textmu eV, and a lever arm with respect to gate $V_\mathrm{PG}$ of $\alpha_\mathrm{PG}=0.021$, similar to that of Device~1. From the height of the leftmost even Coulomb diamond we extracted a single-particle excitation energy of $\Delta\approx170~\mathrm{\mu eV}$, which is significantly lower than the value found for Device~1. All odd Coulomb diamonds in Fig.~\ref{fig:3}(a) feature Kondo peaks and all even Coulomb diamonds feature inelastic cotunneling. Zoom-ins of the Coulomb diamond marked in Fig.~\ref{fig:3}(a) at zero and finite magnetic field are depicted in Fig.~\ref{fig:3}(b) and (c), respectively, and show that the Kondo peak splits in a finite magnetic field. Figure~\ref{fig:3}(d) shows that the Kondo splitting at $V_\mathrm{PG}=-2.344~$V is indeed linear up to 100~mT.

In the following, we use two distinct signatures of the Zeeman splitting at finite magnetic field to determine the $g$-factor, namely Kondo splitting and level splitting between ground and excited state of an unpaired spin at odd electron filling [see the yellow double-arrow in Fig.~\ref{fig:2}(f)]. Both of these energy splittings have been widely used for the extraction of $g$-factors in quantum dots \cite{Jespersen2006,Nilsson2009,dHollosy2013,Mu2020,Fasth2007,Bjork2005,Hanson2003,Escott2010,Hanson2007}. For the Kondo splitting, the effective $g$-factor was extracted as \cite{GoldhaberGordon1998,Cronenwett1998,Inoshita1998,Jespersen2006}:
\begin{equation}
	|g^*|=\frac{e \Delta V_\mathrm{SD}}{2 \mu_\mathrm{B} B},
	\label{eq:1}
\end{equation}
where $\mu_\mathrm{B}$ is the Bohr magneton and $\Delta V_\mathrm{SD}$ is the separation of the two maxima in $G(V_\mathrm{SD})$. For the excited state level splitting, which measures variations of energy levels with respect to only one lead of the device, a pre-factor $(1\pm \delta \alpha)$ needs to be included in Eq.~\ref{eq:1}  \cite{Fasth2007}. The quantity $\delta \alpha = \alpha_\mathrm{S}-\alpha_\mathrm{D}$ is the difference in source and drain lever arm and accounts for the asymmetric coupling to source and drain. For further details on this see the Supplementary Information.

The $g$-factor extracted from the Kondo splitting in Fig.~\ref{fig:3}(c) is 3.8. Repeating the analysis for all the Kondo peaks in Fig.~\ref{fig:3}(a) yields $g$-factors between 0 and 4.8. These results are shown in Supplementary Figure~\ref{fig:S2} and they indicate that the $g$-factor strongly varies with gate voltage. Here, we focus on the state marked with a white arrow in Fig.~\ref{fig:3}(a) and investigate its $g$-factor for different magnetic field orientations. To this end, a magnetic field with a fixed magnitude of 100~mT was rotated by 360\textdegree{} in steps of 15\textdegree{} along three orthogonal planes. Charge stability diagrams such as Fig.~\ref{fig:3}(c) were obtained for all magnetic field orientations and the $g$-factor was extracted from Kondo splittings, because level splittings could not be resolved at all magnetic field orientations. A similar analysis was carried out for both regimes of Device~1, where the out-of-plane rotations were perpendicular and parallel to the nanowire axis. For the latter device, a magnetic field magnitude of 200~mT was used and the $g$-factor was extracted from the level splittings. Schematics of the magnetic field rotations are depicted in Fig.~\ref{fig:S3}(g) and (h) for Device~1 and 2, respectively, together with definitions of the azimuthal angle $\phi$ and polar angle $\theta$, which were used to define the rotations. The set of $g$-factors of each regime was fit with an effective $g$-factor tensor, which describes the $g$-factor as a function of magnetic field direction:
\begin{equation}
	|g^*|(\vec{B})=\frac{1}{|\vec{B}|}\sqrt{g_1^2 B_1^2 + g_2^2 B_2^2 + g_3^2 B_3^2},
	\label{eq:2}
\end{equation}
where $g_\mathrm{i}$ are the principal $g$-factors, pointing along the principal axes of the effective $g$-factor tensor, and $B_\mathrm{i}$ are the magnetic field components along the principal axes \cite{Brouwer2000,Mu2020}.

The results of this analysis are shown in Fig.~\ref{fig:4} with $g$-factors extracted from energy level splittings (Device~1, two regimes) and Kondo splittings (Device~2). Figure~\ref{fig:4}(a-c) depicts polar plots of the $g$-factors extracted for the in-plane rotations of the magnetic field (red) and the fits of the tensor $|g^*|(\vec{B})$ (blue). The nanowires are shown schematically in each polar plot. The polar plots showing the $g$-factors for the out-of-plane magnetic field rotations are presented in Supplementary Figure~\ref{fig:S3}(a-f). Figure~\ref{fig:4}(d-f) shows the extracted $g$-factors for all magnetic field rotations (red, purple, green lines) and the fits of Eq.~\ref{eq:2} with the principal $g$-factors from Table \ref{tab:1} (surface plots) for each regime. The principal $g$-factors are depicted as black lines. The values of the principal $g$-factors, as well as their polar and azimuthal angles, are displayed in Table \ref{tab:1}.\\
The $g$-factor is anisotropic for all investigated regimes in Fig.~\ref{fig:4}. The values of the principal $g$-factors vary between 0.9 and 22.4, depending on the magnetic field orientation. Moreover, the in-plane $g$-factors are typically smaller than the out-of-plane $g$-factors.\\

\begin{table}[]
		\centering
	\caption{\textbf{Principal $g$-factors of the effective $g$-factor tensor from Eq.~\ref{eq:2} for all investigated quantum dot regimes.}}
	\begin{tabular}{cccc}
		\hline
		{\textbf{Device~1, regime 1} } & {$\boldsymbol{g_1}$ } & { $\boldsymbol{g_2}$} & {$\boldsymbol{g_3}$} \\
		\textbf{Value} &  9.3 & 14.1 & 0.9 \\
		$\boldsymbol{\phi}$  & 276.5\textdegree   & 40.8\textdegree & 159.0\textdegree \\ 
		$\boldsymbol{\theta}$ &  53.4\textdegree  & 52.8\textdegree & 58.1\textdegree \\ \hline
		{\textbf{Device~1, regime 2} } & {$\boldsymbol{g_1}$ } & { $\boldsymbol{g_2}$} & {$\boldsymbol{g_3}$} \\
		\textbf{Value} &  7.0 & 22.4 & 3.3 \\ 
		$\boldsymbol{\phi}$  & 144.3\textdegree   & 149.8\textdegree & 54.7\textdegree \\ 
		$\boldsymbol{\theta}$ &  104.9\textdegree  & 14.9\textdegree & 88.6\textdegree \\ \hline 
		{\textbf{Device~2} } & {$\boldsymbol{g_1}$ } & { $\boldsymbol{g_2}$} & {$\boldsymbol{g_3}$} \\
		\textbf{Value} &  5.1 & 11.2 & 2.3 \\ 
		$\boldsymbol{\phi}$  & 187.7\textdegree   & 142.9\textdegree & 95.7\textdegree \\ 
		$\boldsymbol{\theta}$ &  100.8\textdegree  & 15.0\textdegree & 100.3\textdegree \\ \hline
	\end{tabular}
	\label{tab:1}
\end{table}

The experimental results have been presented and will be discussed in the remainder of this paper. PbTe is expected to have an extremely large dielectric constant $\epsilon_{\mathrm{r}}\sim1350$ at low temperatures \cite{Guenther2005}. It is therefore crucial to understand how this value impacts the physics of our quantum dots. We consistently found small charging energies, which might be due to the large dielectric constant of PbTe, which effectively screens the electron cloud within the PbTe wire from the surroundings of the dot. Our results are similar to the observation of vanishingly small charging energies in vertically grown PbTe nanowires \cite{Gomanko2021}, which were also interpreted as consequences of the large dielectric constant of PbTe. However, unlike Ref. \cite{Gomanko2021}, our devices always show finite, albeit small, charging energies. Differences in the properties of the PbTe nanowires and in the quantum dot geometries could explain these dissimilarities.

Due to the large dielectric constant, understanding the impact of side gates is not trivial. The expansion of field lines at the interface between a material with small $\epsilon_{\mathrm{r}}$ (vacuum, SiN${}_\mathrm{x}$ or InP) and one with large $\epsilon_{\mathrm{r}}$ (PbTe) might result in a side gate affecting the chemical potential of the nanowire over a length much larger than the gate width. In this scenario, the quantum dots would not be defined by the side gates but by the length of the entire nanowire (2~\textmu m). The gate lever arms measured in all regimes were similar and approximately equal to 0.01. This indicates that the center of the quantum dot coincides with the center of the nanowire. Since all gate lever arms of a quantum dot need to sum to unity \cite{Ihn2010}, we deduce that $\alpha_\mathrm{S}$ and $\alpha_\mathrm{D}$ are substantially larger than the gate lever arms, as confirmed by the quantitative analysis presented in the Supplementary Information. The length $L$ of the quantum dot can be estimated from the level spacing $\Delta$. We omit the smallest dimension (height) for simplicity and consider the quantum dot as an ellipse with an aspect ratio of 1:10, thus conserving the aspect ratio of the nanowire region between the contacts. Using $\Delta =(1/\pi)\hbar^2\pi^2/\left(m^*(\pi/4) L^2/10\right)$ \cite{Kouwenhoven1997}, where $m^*$ is the effective electron mass $m^*=0.024m_\mathrm{e}-0.24m_\mathrm{e}$ \cite{Ridolfi2015} and $\Delta=170-500~$\textmu eV, we find $L\sim160-860~$nm. This result implies that the quantum dots are likely formed between the contacts, which could indicate that the dielectric constant of nanowires is lower than the expected bulk dielectric constant of PbTe. A reduction of the dielectric constant with decreasing nanowire diameter was already found for ZnO nanowires \cite{Yang2012}. Future investigations of non-local gating and dielectric constant reduction can give more insight into quantum dot formation in PbTe nanowires.\\
The single-particle excitation energy for Device~1, $500~$\textmu eV, was significantly larger than that of Device~2, $170~$\textmu eV, which we attribute to either the different nanowire thicknesses, resulting in a stronger confinement and thus a larger single-particle excitation energy for Device~1, or to different effective masses in the devices. For instance, a large dependence of the electron effective mass on nanowire crystal direction was predicted \cite{Cao2022}. A more systematic study of nanowire geometry and crystal orientation is needed to exclude device to device variability as the root cause of the observed different single-particle excitation energies.\\

Using two distinct methods to extract the $g$-factor, namely Kondo splitting and level splitting, we found strong $g$-factor anisotropies for all investigated regimes. By comparing the level splitting to the Kondo splitting for several magnetic field rotations, we found that the Kondo splitting underestimates extracted $g$-factors by about 20\% compared to the level splitting. This is in qualitative agreement with expectations from theory \cite{Moore2000}. Thus, whenever the excited state level splitting can be resolved, this method should be preferred over Kondo splitting for extracting $g$-factors. The principal axes of the effective $g$-factor tensors $|g^*|(\vec{B})$ for the different regimes are neither aligned, nor perpendicular to the nanowire axes. Moreover, the $g$-factor anisotropy did not present any correlation to the nanowire crystal directions. The $g$-factor anisotropy varied for the two devices and for different gate voltage regimes within Device~1. These observations point to strong Rashba spin-orbit interaction and asymmetric confinement potentials in the PbTe quantum dots \cite{Mu2020}. This is consistent with predicted small Dresselhaus SOI in PbTe due to its inversion-symmetric rocksalt crystalline structure \cite{Peres2011SOI} and large Rashba SOI, as measured in PbTe quantum wells \cite{Peres2011}. Moreover, $g$-factor anisotropy was predicted for [100] and [111] PbTe quantum wells \cite{Ridolfi2015}, where the authors included the contributions of wavefunction barrier penetration, confinement energy shift, and interface SO interaction in their calculations of the quantum well $g$-factors. Furthermore, they found that a confining mesoscopic potential renormalizes the $g$-factor through Rashba SOI.\\
Besides the $g$-factor anisotropy, we observed that the $g$-factor varies for neighboring electronic states in Device~2, similar to results on quantum dots in InAs nanowires \cite{Csonka2008}, where the authors attributed this $g$-factor variation to random fluctuations in the confinement potential. Additionally, the $g$-factors that we found are typically lower than the $g$-factors of bulk PbTe. It is known that the $g$-factor is reduced in low dimensions due to quantum confinement, which leads to quenching of the orbital angular momentum, as observed for quantum dots in InAs nanowires \cite{dHollosy2013}. 

In conclusion, we characterized quantum dot devices in zero and finite magnetic fields. The SAG approach allowed investigation of quantum dots in nanowires with different crystal directions. Despite SAG of PbTe was only recently achieved \cite{Jiang2022}, we could identify regimes with electronic stability that allowed extensive characterization. Charging energies and single-particle excitation energies were extracted from charge stability diagrams. From the energy level and Kondo peak splitting at finite magnetic fields, we extracted the electron $g$-factor as a function of magnetic field direction. The anisotropy of the $g$-factor was attributed to strong Rashba SOI and quantum confinement. Therefore, PbTe in combination with a superconductor is a promising platform for studying topological superconductivity. The large $g$-factor anisotropy and the fact that small $g$-factors are observed for in-plane magnetic fields should be considered in device design.

\section{Methods}
\textbf{Device Fabrication.} Nanowires were selected by imaging with scanning electron microscopy. A double resist layer, consisting of PMMA AR-P 669.04 and PMMA AR-P 672.02, was spun onto the chip and patterned with e-beam lithography. After developing the resist with MIBK:IPA (1:2), an Ar reactive ion etch was performed to remove native oxide on the PbTe nanowires \cite{Yang2008}. Immediately after the Ar etch, the chip was loaded into an e-beam evaporator where 5~nm Ti and 50~nm Au were deposited. Then, lift-off was carried out in acetone.

\textbf{$g$-factor fitting.} By fitting the complete set of $g$-factors extracted for all magnetic field orientations with Eq.~\ref{eq:2}, we determined the principal $g$-factors and the principal axes of the effective $g$-factor tensor $|g^*|(\vec{B})$. The magnetic field components $B_\mathrm{x}$, $B_\mathrm{y}$, $B_\mathrm{z}$ and the measured $g$-factors formed the set of input parameters for the fit. The fit parameters were the principal $g$-factors $g_1$, $g_2$, $g_3$ and the Euler angles of rotation $\phi$, $\theta$, $\psi$ \cite{Liles2021}. With these angles, the magnetic field components were transformed from the Cartesian coordinate system to the coordinate system of the principal axes of $|g^*|(\vec{B})$. This fitting procedure was repeated for two gate voltage regimes of Device~1 and for one regime of Device~2. Subsequently, with the Euler angles of rotation found by the fits, we transformed the principal $g$-factors to spherical coordinates to determine the orientation of each principal $g$-factor.

\section{Acknowledgments}
We thank W.~Riess, G.~Salis, E.~G.~Kelly, F.~J.~Schupp and the Cleanroom Operations Team of the Binnig and Rohrer Nanotechnology Center (BRNC) for their help and support. A.~Fuhrer acknowledges support from NCCR SPIN, funded by the SNSF under grant number 51NF40-180604. The work in Eindhoven is supported by the European Research Council (ERC TOCINA 834290). F.~Nichele acknowledges support from the European Research Council, grant number 804273, and the Swiss National Science Foundation, grant number 200021\_201082.

\bibliography{library.bib}

\clearpage
\newpage
\newcounter{myc} 
\renewcommand{\thefigure}{S.\arabic{myc}}

\section{Supplementary Information}

\subsection{Additional charge stability diagrams}
\setcounter{myc}{1}
\begin{figure}[b]
	\includegraphics[width=\columnwidth]{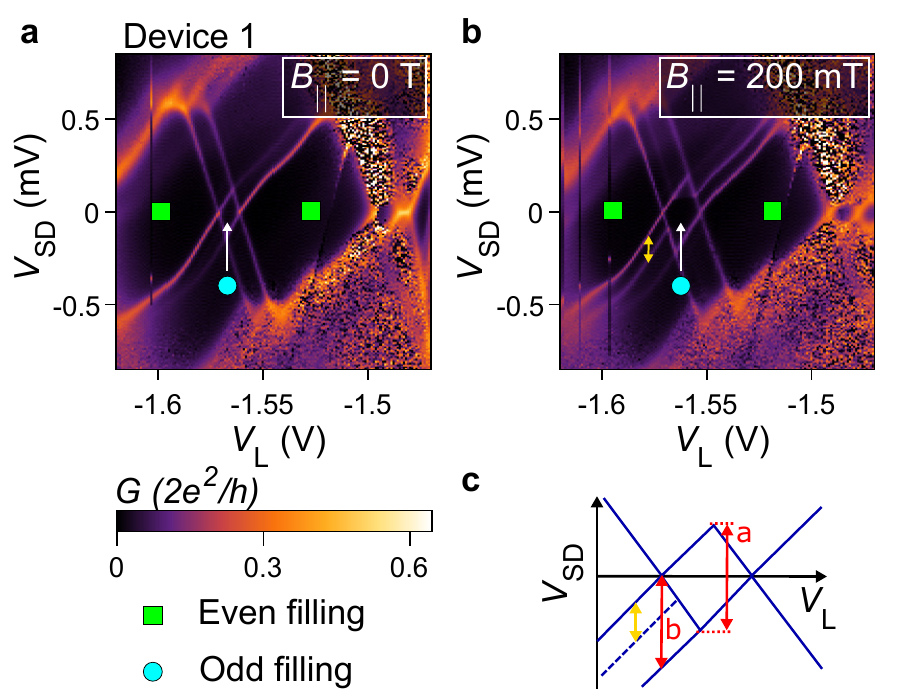}
	\caption{\textbf{Additional measurements of Device~1.} \textbf{a,b} Charge stability diagrams of regime 2 at zero and finite magnetic field. $V_{\mathrm{PG}}=-1,25~$V and $V_\mathrm{R}=-2.4~$V. No Kondo peak is observed in the odd Coulomb diamond. The level splitting is indicated with a yellow double-arrow in (b). \textbf{c} Schematic of an odd Coulomb diamond, showing the level splitting and pre-factor elements $a$ and $b$.}
	\label{fig:S1}
\end{figure}
Charge stability diagrams of Device~1, regime 2 at zero and finite magnetic field are shown in Fig.~\ref{fig:S1}(a) and (b), where $V_{\mathrm{PG}}=-1,25~$V, $V_\mathrm{R}=-2.4~$V, and $\alpha_\mathrm{L}\approx0.012$. Surprisingly, in Fig.~\ref{fig:S1}(a) there is no Kondo peak in the indicated odd Coulomb diamond. It is uncertain whether there is a Kondo peak at $V_\mathrm{L}=-1.48~$V, since what appears to be a resonance of the quantum dot obscures this region of the charge stability diagram. The charging energy was determined from the indicated odd Coulomb diamond: $E_\mathrm{C}\approx120~$\textmu eV. The single-particle excitation energy was extracted from the leftmost even Coulomb diamond: $\Delta\approx460~$\textmu eV. These values are similar to those found for regime 1 of this device. The conductance peaks at the tips of the even Coulomb diamonds are most likely signatures of inelastic cotunneling, yet these regions of the charge stability diagram are ill-defined.\\
\setcounter{myc}{2}
\begin{figure}[t]
	\includegraphics[width=\columnwidth]{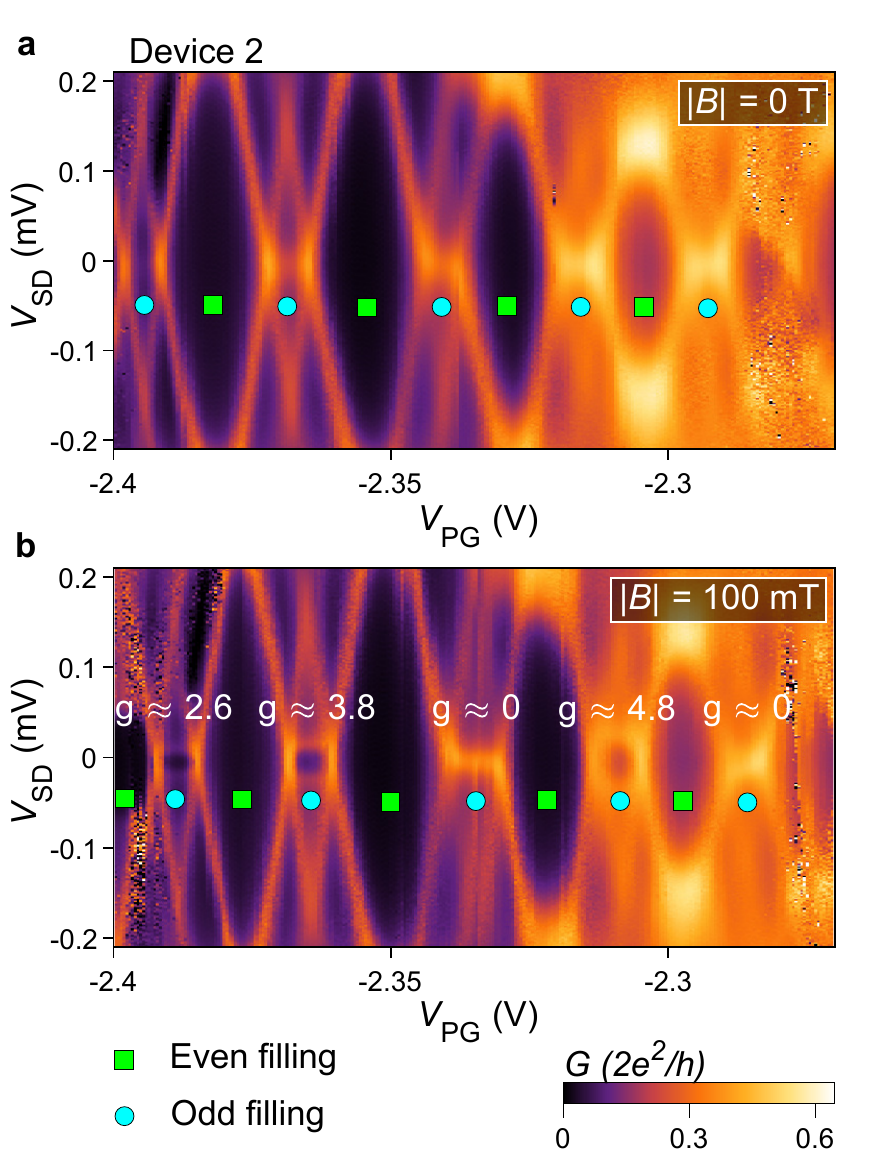}
	\caption{\textbf{Additional measurements of Device~2.} \textbf{a,b} Charge stability diagrams of Device~2 at zero and finite magnetic field. The Kondo peaks in odd Coulomb diamonds show different energy splittings and thus we find different $g$-factors. $V_\mathrm{L}=-3.825~$V and $|B|=100~$mT, with $\phi=-30$\textdegree and $\theta=75$\textdegree{}.}
	\label{fig:S2}
\end{figure}
\setcounter{myc}{3}
\begin{figure*}
	\includegraphics[width=2\columnwidth]{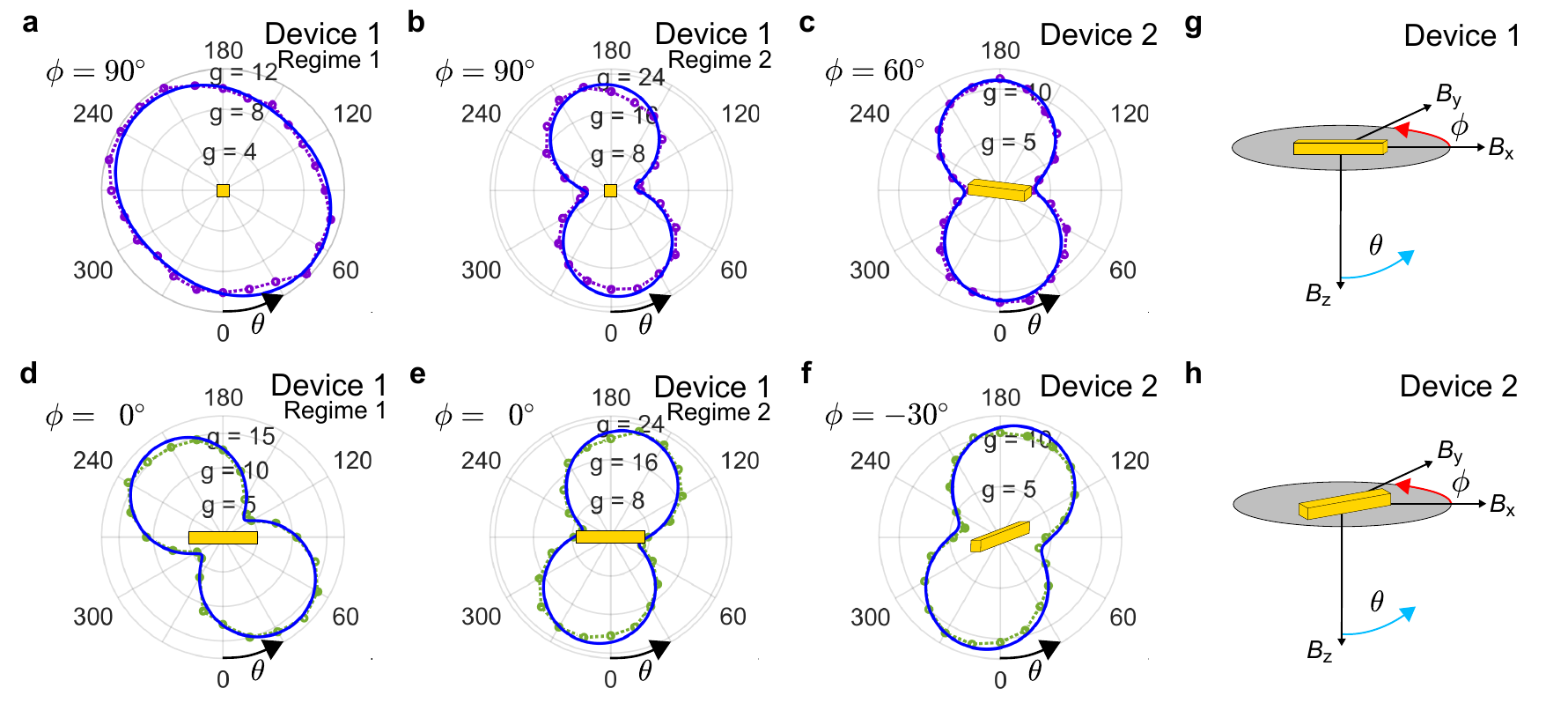}
	\caption{\textbf{Additional polar plots and magnetic field rotation schematics.} \textbf{a-f} Out-of-plane $g$-factors, extracted from level and Kondo peak splittings (purple, green) and fits of the effective $g$-factor tensors (blue). The magnetic field was rotated in steps of 15\textdegree. The nanowires are displayed in each polar plot. \textbf{g,h} Schematics of magnetic field rotations for Device~1 (g) and Device~2 (h). The nanowires devices (orange) and substrate planes (gray) are depicted.}
	\label{fig:S3}
\end{figure*}
The bias voltage was applied anti-symmetrically, therefore we calculate the asymmetry in capacitive coupling to the source and drain contacts $\delta \alpha$ \cite{Ihn2010}. For regime 1 [in Fig.~\ref{fig:2}(e)] we find an average value of $\delta \alpha = \alpha_\mathrm{S}-\alpha_\mathrm{D}\approx-0.27$, indicating a significantly larger coupling to the drain contact. For regime 2 we find $\delta \alpha\approx-0.28$. Since $\delta \alpha$ is negative for both regimes, we conclude that for these measurements, the asymmetry in capacitive coupling to source and drain is independent of the configuration of the side gates. From the slopes of the Coulomb blockade peaks in Fig.~\ref{fig:2}(a-d), we deduce that the lever arms of gates $V_\mathrm{L}$ and $V_\mathrm{R}$ are similar for Device~1. From additional measurements, we found that the gate lever arm of $V_\mathrm{PG}$ is larger, but within the same order of magnitude. The sum of source, drain and all gate lever arms of a quantum dot is equal to 1 \cite{Ihn2010}. Since the sum of the gate lever arms does not exceed 0.1, approximately 90\% of the total lever arm, and therefore the capacitance of the quantum dot, can be attributed to the source and drain coupling. This can be understood in the framework of the device design, Fig.~\ref{fig:1}(a): the contacts are wrapped around the nanowire, while the gates are positioned at a distance of $\sim80~$nm away from the nanowire, resulting in a large capacitive coupling to the source and drain and a small capacitive coupling to the gates. From $\delta \alpha \approx-0.28$ we can estimate the source and drain lever arms, yielding $\alpha_\mathrm{S}\approx 0.31$ and $\alpha_\mathrm{D}\approx0.59$. For Device~2, the average asymmetry in capacitive coupling is $\delta \alpha \approx0.21$ and again, we found that both gate lever arms are in the order of a few percent. Thus, also for this device the source and drain lever arms dominate the capacitance of the quantum dot.\\

At finite magnetic field, we observe spin excited states of the even Coulomb diamonds around the indicated odd Coulomb diamond in Fig.~\ref{fig:S1}(b). The $g$-factor was extracted from the energy level splitting between the ground state and the corresponding excited state, indicated with a yellow double-arrow. In general, this energy splitting is equal to twice the Zeeman splitting \cite{Escott2010,Hanson2007}. However, the Coulomb diamonds are skewed and therefore, the level splitting must be multiplied with a pre-factor. This pre-factor is equal to $a/b$, where $a$ is the energy difference between the tips of the odd Coulomb diamond and $b$ is the energy difference between the zero energy crossing of the odd Coulomb diamond edge and the extension of the Coulomb diamond boundary. Both $a$ and $b$ are indicated with red arrows in Fig.~\ref{fig:S1}(c). In this schematic, $a/b>1$. Moreover, the pre-factor can also be expressed in terms of the asymmetry in source and drain lever arms: $a/b=(1\pm \delta \alpha)$ \cite{Fasth2007}. If the Coulomb diamond is symmetric, i.e., if $\delta \alpha =0$, then $a/b=1$. For the Kondo peak splitting the pre-factor is not easily taken into account since both leads play a role in forming this highly correlated state. We expect that the Kondo splitting reflects an average between the two pre-factors for each lead and is thus close to $\delta \alpha=0$ (pre-factor equals 1). However, the extraction of energy splittings from the maxima of split Kondo peaks has been shown to underestimate the Zeeman splitting \cite{Moore2000} and indeed we find that the $g$-factors extracted from Kondo splittings are approximately 20\% lower than those extracted from level splittings in the same charge stability diagrams. 

Charge stability diagrams of Device~2 at zero and finite magnetic field are depicted in Fig.~\ref{fig:S2}. These charge stability diagrams were obtained in the same regime as Fig.~\ref{fig:3}(a), however with slightly different voltage ranges for $V_\mathrm{PG}$ and $V_\mathrm{SD}$. By comparing the charge stability diagrams in Fig.~\ref{fig:3}(a) and \ref{fig:S2}(a), which were measured several days apart, it is clear that the quantum dot energy levels shift over time. However, the two charge stability diagrams in Fig.~\ref{fig:S2} were measured within a time frame of hours. A shift of the energy levels can be seen between Fig.~\ref{fig:S2}(a) and (b), however this shift is inconsequential for the $g$-factor measurements. For Fig.~\ref{fig:S2}(b), $|B|=100~$mT, with the azimuthal angle $\phi$ equal to $-30$\textdegree{} and the polar angle $\theta$ set to 75\textdegree. At this orientation of the magnetic field, the Kondo peaks show significantly different energy splittings. This observation indicates that the $g$-factor varies with gate voltage, which was also found in a study on quantum dots in InAs nanowires \cite{Csonka2008}.

\subsection{Additional polar plots}
In Fig.~\ref{fig:S3}(a-f), polar plots of the out-of-plane magnetic field rotations are depicted: (a-c) show the out-of-plane rotations perpendicular to the nanowire axis for Device~1 and 30\textdegree{} off-axis for Device~2, and (d-f) show the out-of-plane rotations parallel to the nanowire axis for Device~1 and 60\textdegree{} off-axis for Device~2. Schematics of the nanowires are included in the polar plots. In Fig.~\ref{fig:S3}(g) and (h), the methodology for rotating the magnetic field is depicted schematically for both quantum dot devices. The in-plane rotations were performed by increasing the azimuthal angle $\phi$ from 0\textdegree{} to 360\textdegree{} in steps of 15\textdegree, with the polar angle $\theta$ set to 0\textdegree. The out-of-plane rotations were performed by setting $\phi$ to a fixed value and rotating $\theta$.

\end{document}